# Do job seekers value procedure in AI hiring only for error correction? Evidence from a conjoint experiment

Chuyao Wang (a, b), Patrick Sturgis (a), and Daniel de Kadt (c)

(a) Department of Methodology, London School of Economics and Political Science, London, United Kingdom

(b) Data Science Institute, London School of Economics and Political Science, London, United Kingdom

(c) Cornell University, Ithaca, New York, United States

Corresponding author: Chuyao Wang, c.wang85@lse.ac.uk.

## Abstract

Employers increasingly delegate initial screening to automated systems, which in many cases reject an application before any human reads it. Acceptance of such systems plausibly depends both on how well they perform and on the procedure that produces the decision. Prior studies rarely vary procedure and performance independently, leaving it unclear whether applicants value procedure for its own sake or for the errors it corrects. In a preregistered paired-profile conjoint experiment, 1,919 United States job seekers made eight choices between systems with independently randomized levels of decision authority, error rate, explanation, opt-out, appeal, and independent bias audit. The value of the appeal, the opt-out, and the bias audit did not rise as wrongful rejections became more common, each staying within a preregistered equivalence bound. Human involvement carried more weight than any procedural feature, moving stated choice about as much as cutting wrongful rejections from 30% to 10%. These patterns constrain a simple error-correction account and are consistent with applicants valuing procedure partly for its own sake, so that improving a system's performance does not substitute for a right applicants can invoke.

## 1 Introduction

Employers increasingly delegate part of hiring to AI systems. In surveys of United States human resources professionals, the share of organizations using AI for HR tasks rose from 26% in 2024 to 43% in 2025 (Society for Human Resource Management, 2025). Hiring is the most common area of use (Society for Human Resource Management, 2022). In a survey of executives in the United States, the United Kingdom, and Germany, more than 90% of those with a recruitment management system used it to filter or rank applicants, and 88% agreed that qualified candidates are screened out because their resumes do not match the job description exactly (Fuller et al., 2021).

Public attitudes remain cautious. Most Americans oppose AI making final hiring decisions, and about two-thirds would not want to apply to an employer that uses AI to help make them (Pew Research Center, 2023). Nationally representative evidence likewise finds broad doubts about the fairness and effectiveness of hiring algorithms (Zhang & Yencha, 2022), and more recent data show that workers remain more apprehensive than hopeful about workplace AI (Pew Research Center, 2025). Adoption is therefore widespread despite public reservations, which raises the question this study addresses: what makes an AI hiring system more or less acceptable to the applicants it judges, and does the answer depend on how well the system performs?

Employers adopt automated screening to process large applicant pools faster (Fuller et al., 2021; Raghavan et al., 2020). A system can perform well and still lose users, because people abandon an algorithm once they see it err (Dietvorst et al., 2015). Applicants may also value who decides and what they can do about a rejection. If so, a right decision is not enough for employers, and performance standards alone are not enough for regulators. Whether procedure carries weight of its own is what this study tests.

Automated screening is already contested in litigation and in regulation. A federal age-discrimination claim over automatic rejection settled in 2023, and a civil-rights complaint filed in 2024 alleges race and disability discrimination in vendor assessments (American Civil Liberties Union, 2024; U.S. Equal Employment Opportunity Commission, 2023). New York City's Local Law 144 mandates bias audits and candidate notice for covered hiring tools (New York City Council, 2021). The General Data Protection Regulation restricts certain solely automated decisions (European Parliament & Council of the European Union, 2016), and the Artificial Intelligence Act classifies specified recruitment systems as high-risk (European Parliament & Council of the European Union, 2024). Section 5 sets out what each requires.

People evaluate consequential decisions by both their outcomes and the procedures that produce them. Procedural justice research supplies two accounts of why applicants value procedure. One ties that value to the mistakes a feature can detect or reverse, the other to what a procedure communicates about standing and respect (Lind & Tyler, 1988; Thibaut & Walker, 1975; Tyler & Lind, 1992). They diverge in whether that value should grow as wrongful rejections increase. Research on algorithmic decision-making likewise finds that perceived fairness depends on more than accuracy alone, though the weight of procedure relative to performance varies across settings (Lee, 2018; Starke et al., 2022). Neither account can be read off a design that holds performance fixed.

Earlier studies of applicants and other decision subjects have mostly measured fairness perceptions (Lavanchy et al., 2023; Yurrita et al., 2023). By contrast, the main conjoint evidence on preferences over human versus algorithmic decision-makers comes from third parties evaluating decisions about others (Bansak & Paulson, 2024), and those targeted by a decision react differently from observers (Langer & Landers, 2021). Here the respondent is the person the system judges.

The study reports a preregistered paired-profile conjoint experiment, drawing on 1,919 United States job seekers who each made eight forced choices between systems that varied decision authority, the error rate, explanation, opt-out, appeal, and independent bias audit. It combines randomized conjoint attributes with preregistered equivalence tests, so that a nonsignificant interaction can be distinguished from evidence that moderation falls within a prespecified equivalence bound.

## 2 Relevant literature and hypotheses

Applicants may value procedure in AI hiring (Gilliland, 1993; Hausknecht et al., 2004), over and above the performance of the system. The pre-analysis plan designated four attributes for confirmatory testing: human involvement, appeal, the opt-out, and the bias audit. An error-correction account treats the confirmatory attributes as instrumental, valued for their capacity to detect or reverse mistakes. A performance-independent account holds that they can retain value regardless of how accurate the system is.

Distinguishing the two accounts requires varying who decides, how often the system errs, and what recourse an applicant has independently of one another. Explanation was measured and is reported separately. Appeal, the opt-out, explanation, and the bias audit are referred to throughout as procedural features.

### 2.1 Why human involvement may matter more than procedural features

Prior studies have tested the value of these attributes, but each holds performance fixed or places the respondent outside the decision. Applicants rate fully algorithmic hiring as less fair than human-involved hiring, with performance held fixed (Lavanchy et al., 2023). Decision subjects value contestability and explanations in a loan scenario, again at fixed performance (Yurrita et al., 2023), and rejected borrowers weigh the type and speed of review above the reviewer's identity (Lyons et al., 2022). Where performance is randomized, respondents judge decisions about other people (Bansak & Paulson, 2024) or rate interpretability across many applications from the outside (Nussberger et al., 2022).

Human involvement is distinct because it changes who retains final authority over the decision, whereas appeal, opt-out, explanation, and auditing modify an otherwise automated process. The value applicants place on human involvement may reflect both the relational standing it conveys (Tyler & Lind, 1992) and the concern that automated evaluation cannot weigh an applicant's individual circumstances (Chun et al., 2024). Prior studies find that people judge human-led or human-assisted decisions fairer than fully automated ones (Kern et al., 2022). Preferences between human and algorithmic decision-makers can nonetheless depend on performance and context (Bansak & Paulson, 2024).

> **H1a.** Human involvement, appeal, the opt-out, and the bias audit each raise the probability that a system is chosen.
>
> **H1b**. Human involvement has the largest effect among the four confirmatory attributes.

### 2.2 The interaction between the confirmatory attributes and the error rate

Whether the value of the confirmatory attributes rises with the error rate distinguishes the two accounts. On an error-correction account, applicants value these attributes instrumentally, for the control they confer over the outcome (Thibaut & Walker, 1975). If corrective capacity is what applicants value, attributes that enable correction should gain value as wrongful rejections become more common.

The prediction is sharpest for appeal, which a rejected applicant can invoke directly to obtain human re-review. It is weaker for the other three: human involvement bundles correction with the authority and standing it also conveys; the opt-out leads to a human alternative whose accuracy was never stated; and a bias audit addresses the distribution of errors across groups rather than the overall rate. Explanation, reported separately, does not by itself reverse a decision.

> **H2a (error correction)**. The value of the four confirmatory attributes increases as the error rate rises.

A performance-independent account instead holds that procedures can retain value across levels of system performance. Voice raises fairness judgments even when it comes too late to change the decision (Lind et al., 1990; Tyler, 2006), and in personnel selection procedural fairness predicts applicant reactions (Acikgoz et al., 2020; Gilliland, 1993; Hausknecht et al., 2004), including in automated selection (Ochmann et al., 2024). People are far more willing to use an imperfect algorithm they can adjust, and that preference persists when the permitted adjustment is too small to improve the forecast appreciably, which suggests a desire for some control rather than for more of it (Dietvorst et al., 2018). More recent evidence finds that people value the interpretability of AI decisions at every tested level of system accuracy, although its rated importance fell as accuracy rose (Nussberger et al., 2022). On this account, applicants value procedural features and human involvement partly for what they express about standing and fair treatment, so the effect of each attribute should not increase as the error rate rises.

> **H2b (performance independence)**. The value of the four confirmatory attributes does not increase as the error rate rises, remaining within a prespecified equivalence bound.

### 2.3 Why appeal and the opt-out gain from a human

Appeal supplies a further human re-review, while the opt-out supplies access to a human from the outset. In the terms of exit and voice (Hirschman, 1970), appeal is a form of voice and the opt-out a form of exit. Voice and recourse are more consequential when a human authority can act on them (Lind & Tyler, 1988; Tyler, 2006), so appeal and the opt-out should both gain value when a human is already involved. They could instead substitute for a human when the AI decides alone.

> **H3. Appeal and the opt-out are valued more** when a human is already involved than when AI decides alone.

### 2.4 Why a system-level audit ranks below individual attributes

Appeal and the opt-out provide routes that an applicant can invoke, and human involvement changes who holds decision authority, whereas a bias audit operates at the system level and offers no individual re-review. Research on fair process locates much of the value of a procedure in the voice and recourse it grants a person over the decision that affects them (Lind et al., 1990; Tyler & Lind, 1992). Studies of algorithmic decisions likewise show that affected parties judge them in individual terms such as dignity, contestability, and the wish not to be reduced to a data point (Binns et al., 2018; Chun et al., 2024; Yurrita et al., 2023). A bias audit instead targets group-level disparate impact, so its benefit is diffuse and shared. Eliminating group bias is not always perceived as fair when it overlooks individual circumstances (Newman et al., 2020).

> **H4.** An independent bias audit is valued less than human involvement, appeal, or the opt-out.

## 3 Data and methods

### 3.1 Sample recruitment, ethics, and data quality

The study received ethical approval from the Department of Methodology at the London School of Economics and Political Science on May 28, 2026. The pre-analysis plan was posted publicly on OSF on June 17, 2026 (https://osf.io/5ju4d/), before the data in this study were examined. Registered hypotheses, exclusions, and models follow it, and Appendix G reports the power analysis. Supplementary analyses are identified in the text and summarized in Appendix I. Replication data and code are available at https://github.com/chuyao-wang/ai-hiring-replication.

Respondents were recruited through Prolific and prescreened for United States residence and current job seeking, yielding 1,919 participants; Appendix B reports sample characteristics. All participants provided informed consent and were compensated at a mean of £0.84, approximately £10.22 per hour. Of the respondents, 92% answered the manipulation check correctly, identifying that the error rate varied across tasks. Following the pre-analysis plan, the primary analysis does not condition on completion time, response patterns, or the post-task manipulation check.

### 3.2 Conjoint experimental design and instrument

The instrument is a paired-profile, choice-based conjoint conducted from the applicant's perspective. In each of eight tasks, two hiring systems were described on the six attributes in Table 1. Respondents made a binary forced choice and rated the acceptability of each system on a five-point scale. Appendix A records the additional attitude measures, and the complete wording and materials are available in the OSF repository.

Decision authority took three levels: the AI decided alone, the AI screened applicants before a human decided among those who passed, or the AI made suggestions to a human who decided. The two levels with a human are referred to as human involvement. The error rate was the share of qualified applicants wrongly rejected, set at 10%, 20%, or 30%. Four procedural features were binary: a meaningful explanation of the decision, an option to request a human evaluation from the outset (the opt-out), human re-review after a rejection (appeal), and an independent bias audit.

Table 1: Attributes, with levels as shown to respondents.

| Attribute | Levels (as shown on screen) |
|---|---|
| **Decision authority** | AI suggests, human decides / AI screens applicants first, human decides / AI decides alone |
| **Error rate** | Wrongly rejects about 10% of qualified applicants / about 20% / about 30% |
| **Explanation** | No explanation given / Meaningful explanation |
| **Human option from start (opt-out)** | Not available / Available |
| **Human re-review after rejection (appeal)** | Not available / Available |
| **Independent bias audit** | Not available / Available |

*Note:* Reference levels: "AI decides alone" (decision authority), "no explanation given" (explanation), and "not available" (each binary feature). The error rate is coded continuously in the primary analysis (10%, 20%, or 30% error). Attribute-row order was randomized by task and held fixed across the two profiles. Task position (1–8) was recorded.

Attribute levels were independently and uniformly randomized, subject to the paired systems differing on at least one attribute. Under the standard assumptions for paired-profile conjoint designs (Hainmueller et al., 2014), this randomization identifies the average marginal component effects (AMCEs), the mean change in choice probability when one level replaces the reference. The analytic names used here differ from the on-screen labels. Decision authority was shown as “Decision process”, the error rate as “Accuracy (qualified applicants wrongly rejected)”, and explanation as “Transparency” (Table 2).

Table 2: An example paired-profile task as shown to respondents.

Comparison 8 of 8

| Feature | System A | System B |
|---|---|---|
| Decision process | AI decides alone | AI suggests, human decides |
| Accuracy | Wrongly rejects about 10% of qualified applicants | Wrongly rejects about 30% of qualified applicants |
| Transparency | No explanation given | Meaningful explanation |
| Human option from start | Available | Available |
| Human re-review after rejection | Available | Available |
| Independent bias audit | Not available | Not available |

*Which system would you prefer to be evaluated by?

- System A
- System B

*How acceptable is each hiring system to you?

| | Completely unacceptable | Somewhat unacceptable | Neutral | Somewhat acceptable | Completely acceptable |
|---|---|---|---|---|---|
| System A | ○ | ○ | ○ | ○ | ○ |
| System B | ○ | ○ | ○ | ○ | ○ |

### 3.3 Estimation and inference

The primary estimator is an ordinary least squares regression of the binary profile choice on the randomized attribute levels, with standard errors clustered by respondent (Hainmueller et al., 2014). Two quantities are reported: AMCEs, defined relative to the stated reference levels, and marginal means, the probability that a profile with a given level is chosen, averaging over the other attributes (Leeper et al., 2020). Neither is a majority preference. The AMCE averages over both the direction and the intensity of individual preferences and can therefore differ in sign from the preference of the majority (Abramson et al., 2022). Differences between attribute effects place those effects on a common scale and are not marginal rates of substitution. Interactions between randomized attributes identify causal interaction effects under the same design assumptions, subject to the dependence on reference levels that Egami and Imai (2019) document.

H2 uses a preregistered equivalence test, because a nonsignificant interaction is not evidence of negligible moderation. For each of the four confirmatory attributes, moderation across the 10% to 30% error range is assessed by two one-sided tests at $\alpha = 0.05$ against a bound of $\pm 0.05$ on the choice-probability scale, equivalently by asking whether the 90% confidence interval lies inside that bound (Lakens et al., 2018; Rainey, 2014). The bound was prespecified as the largest change in an attribute effect that would still count as substantively negligible. Because that is a substantive judgment rather than a statistical one, Section 4.5 and Appendix D report how the conclusions change under stricter and wider bounds.

Several checks assess whether these inferences depend on the design or the specification. Six attributes and eight tasks sit within the ranges where prior conjoint work finds no satisficing (Bansak et al., 2018, 2021). The five-point acceptability outcome tests whether the results depend on the forced-choice format, and task position, attribute-row order, and profile position test for ordering and fatigue effects. Logit estimates provide a check on the linear probability results (Ai & Norton, 2003; Mize, 2019). Restrictions based on completion time, response patterns, and the post-task manipulation item test whether the results depend on those diagnostics. Section 4.5 and Appendices E and F report these checks.

## 4 Results

The central result is that the value of the procedural features held steady as the system erred more often. Within the tested range, preregistered equivalence tests bound the change within $\pm 0.05$ on the choice-probability scale. Human involvement fell outside that bound.

### 4.1 Human involvement outweighs the procedural features

Human involvement, appeal, the opt-out, and the bias audit each raised the probability that a profile was chosen (Figure 1), supporting H1a. Relative to AI deciding alone, AI suggestions followed by a human decision raised choice probability by 0.287 and AI screening followed by a human decision by 0.257 (Appendix Table C1). Appeal raised it by 0.156, the opt-out by 0.129, and the bias audit by 0.068. The pooled human involvement effect (the two human-involved levels combined against AI deciding alone) was larger than the effects of appeal, the opt-out, and the bias audit (all $p < 0.001$), supporting H1b. Human involvement, appeal, and the opt-out each exceeded the system-level bias audit (all $p < 0.001$), supporting H4 (Appendix Table C2). Explanation raised choice probability by 0.128 in the prespecified separate analysis.

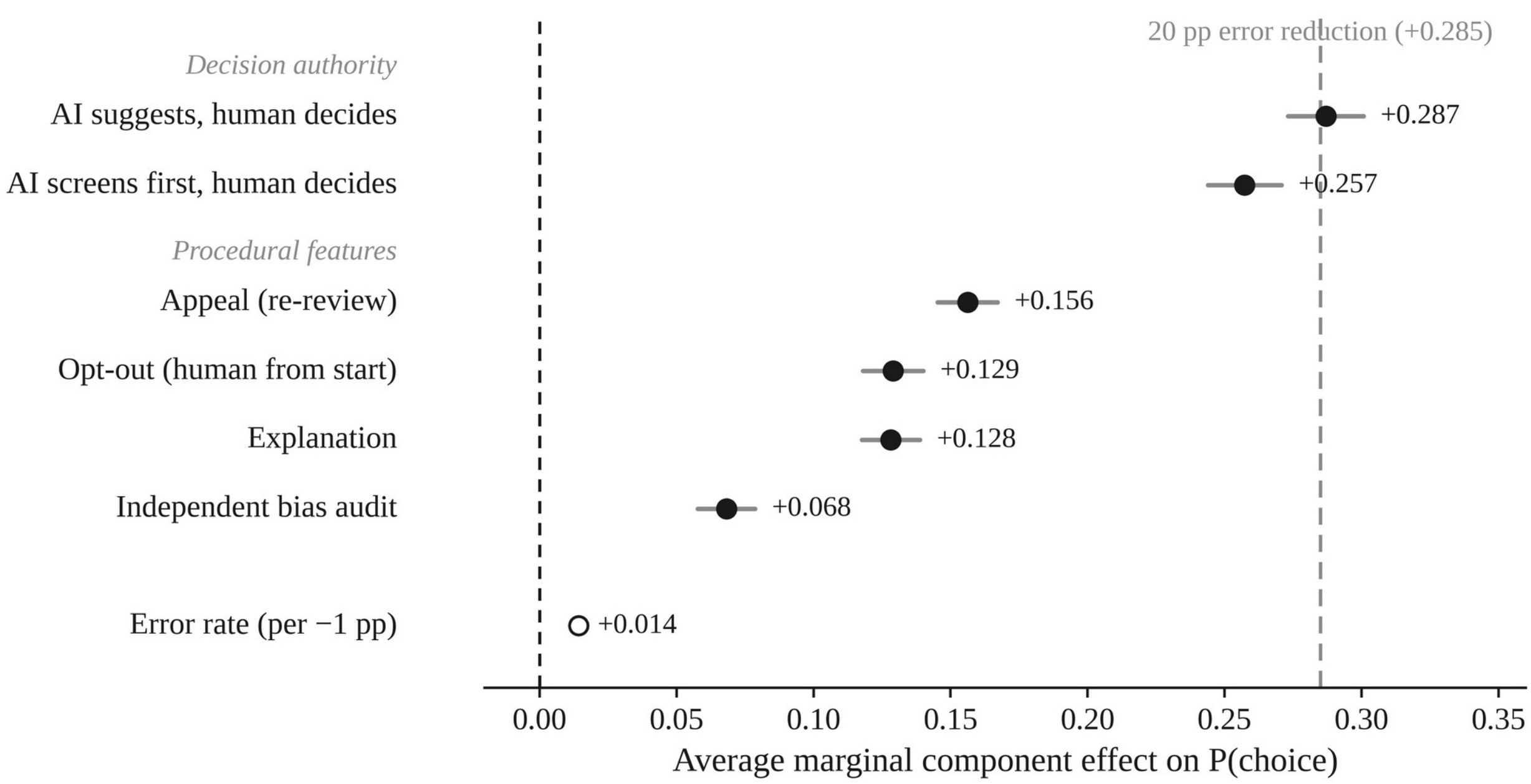

Figure 1: Average marginal component effects on choice probability.

Note: Bars show 95% confidence intervals. The error rate is displayed as the effect of reducing error by one percentage point. The dashed comparison shows the modeled effect of reducing error from 30% to 10%.

Respondents preferred a human decision after AI suggestions to one after AI screening by 0.030 ($p < 0.001$), consistent with evidence that procedural justice falls when AI is granted more decision power than its perceived ability warrants (Jiang et al., 2023). Section 4.2 returns to this contrast.

Reducing the error rate from 30% to 10% raised choice probability by 0.285, the benchmark for the other attribute effects. That is nearly equal to the pooled effect of adding human involvement (0.272): the difference is not distinguishable from zero (95% CI [−0.007, 0.033]).

### 4.2 The value of the procedural features holds as errors rise

At the lowest tested error rate (10%), all four confirmatory attributes had positive conditional effects with confidence intervals excluding zero, satisfying the registered conditional-effect test (Appendix Table I1). Explanation did the same. Because a 10% error rate is not error-free, these conditional effects do not by themselves separate the accounts.

Appeal offers the clearest test of H2, because its corrective value, on an error-correction account, should rise directly with wrongful rejections. Its interaction with the error rate was nonsignificant, with a 90% confidence interval inside the preregistered ±0.05 bound, passing even a stricter ±0.03 bound. The attribute told respondents that re-review was available and said nothing about how well it worked. Section 5 sets out the mechanisms this design cannot exclude.

The opt-out and the bias audit held their value as errors rose and passed the equivalence test, the opt-out at the stricter ±0.03 bound, though neither bears on the overall error rate as directly as appeal. Explanation showed the same pattern. The error-correction prediction (H2a) is therefore supported for none of the attributes, and the performance-independent criterion (H2b) is met for appeal, the opt-out, and the bias audit but not for human involvement.

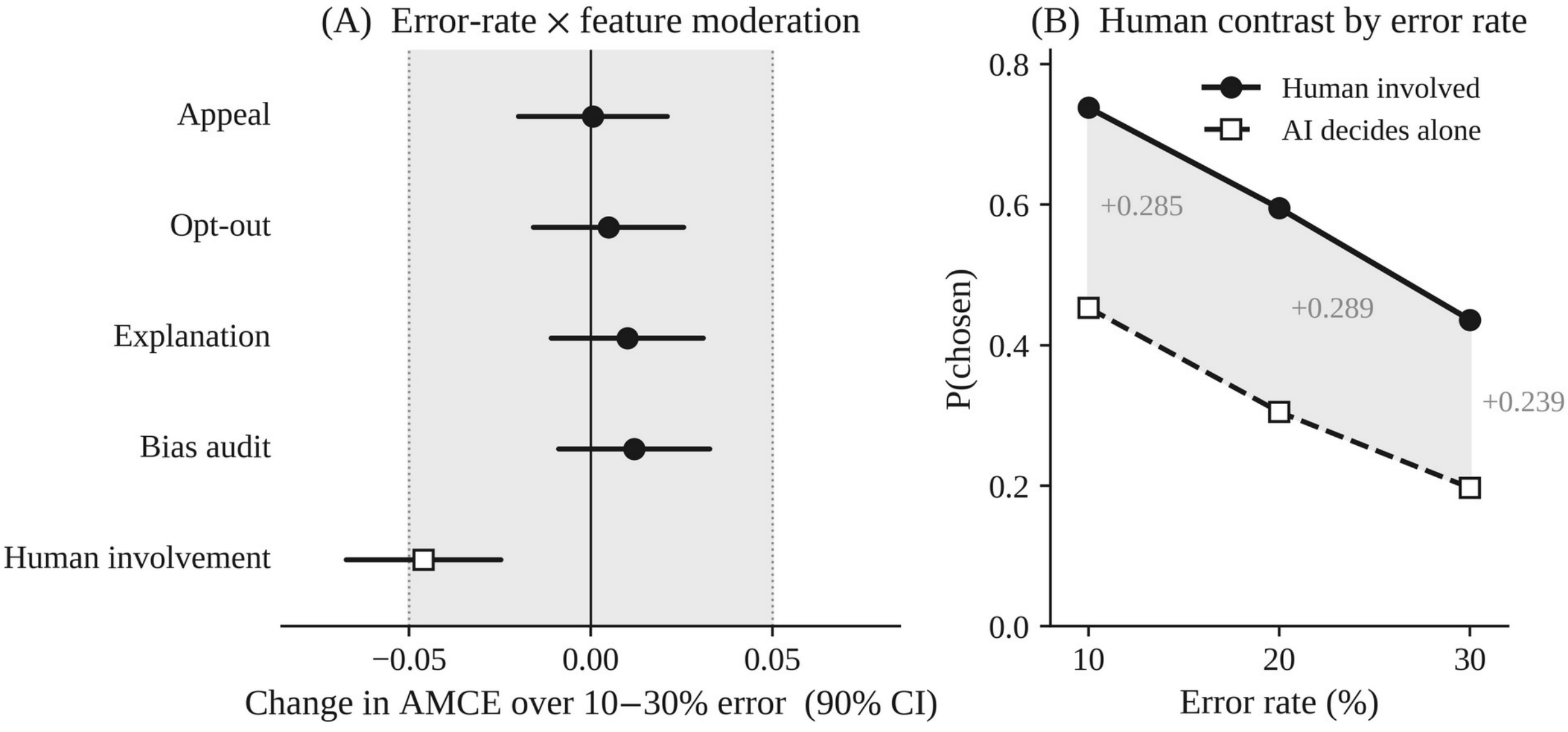

Figure 2: Moderation of the attribute effects by the error rate.

Note: Panel A shows moderation of each attribute effect across the 10% to 30% error range, with 90% confidence intervals and the preregistered equivalence region (±0.05). Explanation is shown from its prespecified separate analysis. Panel B shows the marginal probability of choosing a human-involved versus AI-alone system at each error rate.

Human involvement is the exception (Figure 2, Panel B). Its advantage over an AI-alone system was 0.285 at a 10% error rate, 0.289 at 20%, and 0.239 at 30%, little changed between the first two levels and smaller only at the highest. Under the continuous coding, the registered interaction equals the 30-versus-10 endpoint contrast (−0.046; Appendix Table E2). Its sign is opposite to the error-correction prediction, while the logit interaction is not significant (p =

0.21). The estimate is scale-sensitive and the pattern is non-monotone, so it does not show that human involvement loses value as errors rise.

In an exploratory comparison, applicants preferred “AI suggests, human decides” over “AI screens applicants first, human decides” by a margin that did not move with the error rate (change across the range +0.005, 95% CI [−0.026, 0.036]), although only the former lets a human see the applications the AI would reject.

### 4.3 How appeal and the opt-out combine with human involvement

In the registered two-interaction model, appeal was valued more when a human was already involved than when AI decided alone (+0.025, 95% CI [0.004, 0.047]), supporting its component of H3 on the linear probability scale (Figure 3). Under logit, the same interaction in the joint model is not significant (p = 0.31; Appendix Table E2, Panel B), the same scale sensitivity found for the human-by-error-rate result in Section 4.2. The opt-out interaction was effectively zero (−0.0003; post hoc 90% CI [−0.018, 0.018], within ±0.05), so H3 receives only partial support. The design does not identify why the opt-out interaction was near zero.

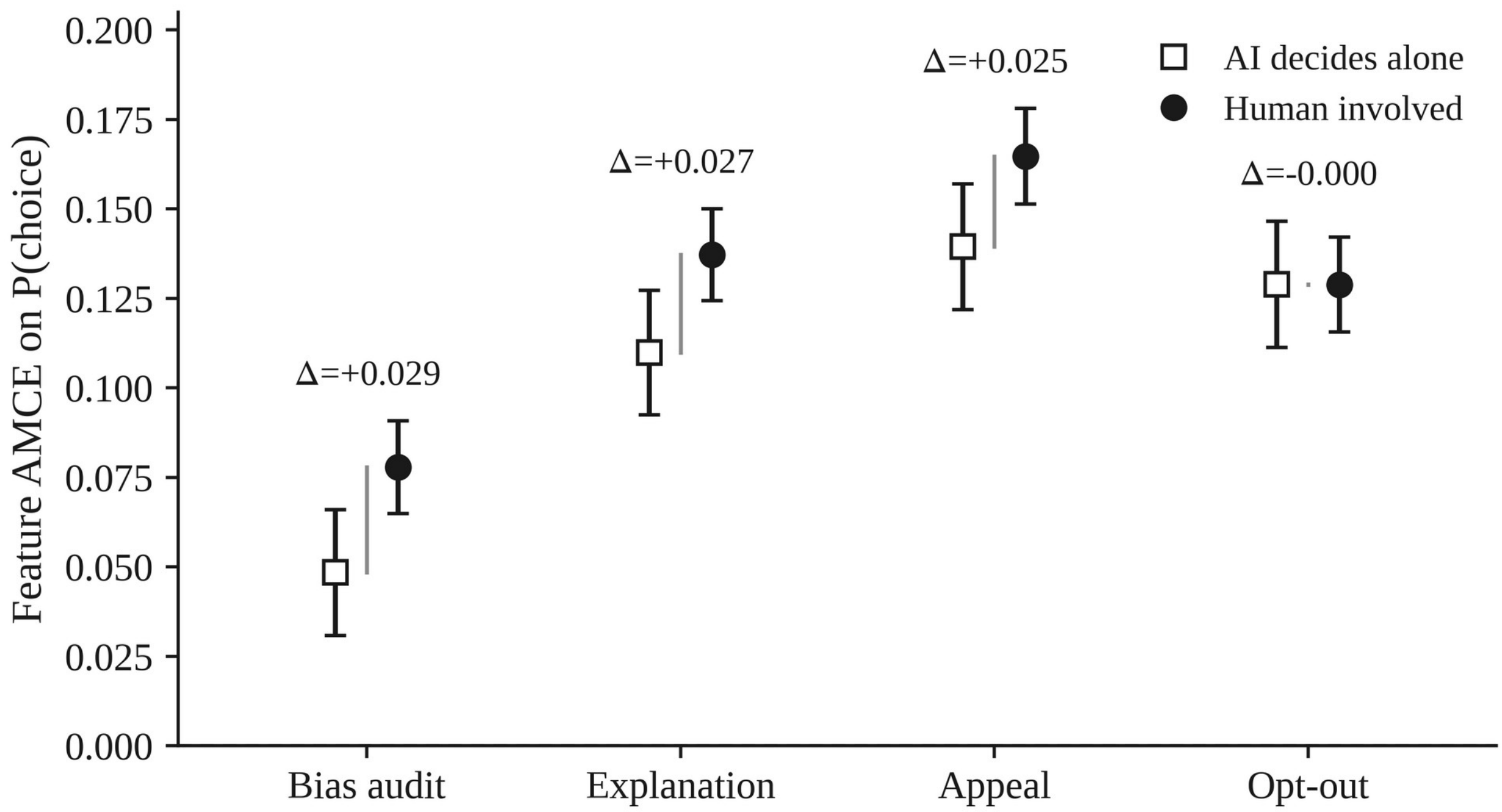

Figure 3: Conditional AMCEs of each procedural feature by human involvement.

Note: Estimates compare profiles where a human is involved with those where AI decides alone; bars show 95% confidence intervals. Only the appeal and opt-out interactions were preregistered. The full set is exploratory.

In an exploratory joint model, appeal, the bias audit, and explanation each gained slightly when a human was involved on the linear probability scale, but only the bias-audit interaction was significant under logit (Appendix Table E2, Panel B).

### 4.4 Legitimacy and intention to apply

Intention to apply to employers using AI hiring averaged 3.24 on the five-point scale, above belief in the legitimacy of AI hiring at 2.85 ($p < 0.001$), a preregistered comparison (Appendix I). The two items measure distinct constructs, a normative judgment and a behavioral intention, and are strongly associated ($r = 0.65$). Among respondents, 7.8% combined below-midpoint legitimacy with above-midpoint intention to apply (Figure 4). Intention to withdraw correlated negatively with both measures (Appendix A).

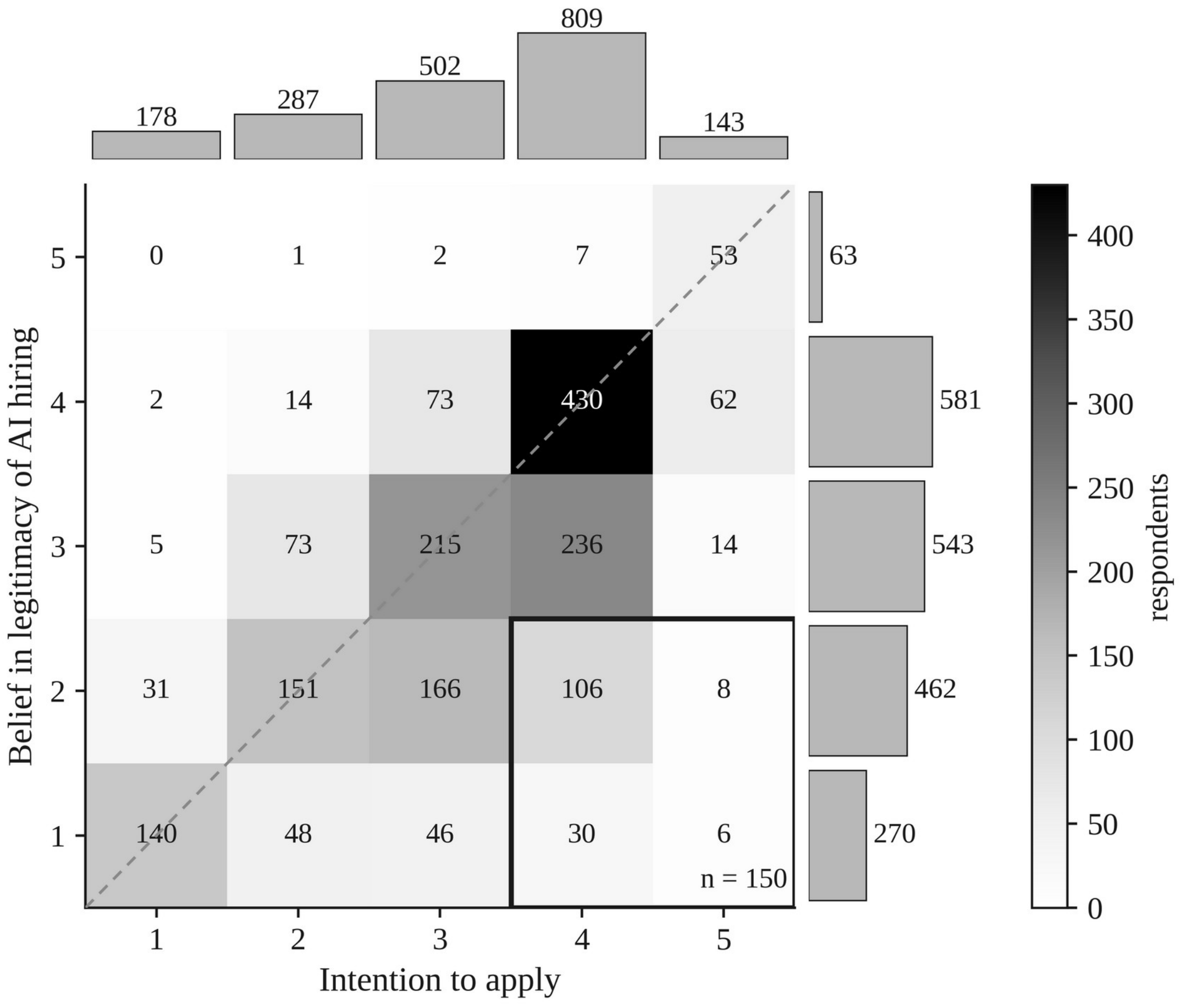

Figure 4: Belief in the legitimacy of AI hiring against intention to apply.

Note: Cells show respondent counts. The dashed diagonal marks equal ratings. The outlined block marks respondents who doubt the legitimacy of AI hiring yet intend to apply (legitimacy ≤ 2 and intention to apply ≥ 4), 7.8% of the sample.

### 4.5 Robustness to specification and scale

Applied one at a time, no data-quality restriction, positional control, or unordered coding of the error rate moves an AMCE by more than 0.003 in choice probability (Appendix F). Exploratory analyses of heterogeneity in the human involvement preference, by prior AI-evaluation experience and by age and gender, provide limited evidence of systematic variation (Appendix H). Neither linear ordering of the three decision-authority levels fits better than treating them as unordered. Soft-launch and remaining batches agree within sampling error, and the five-point acceptability outcome yields the same signs and a similar ordering (rank correlation 0.89). Appendix D and Figure 5 report how the equivalence conclusions depend on the bound.

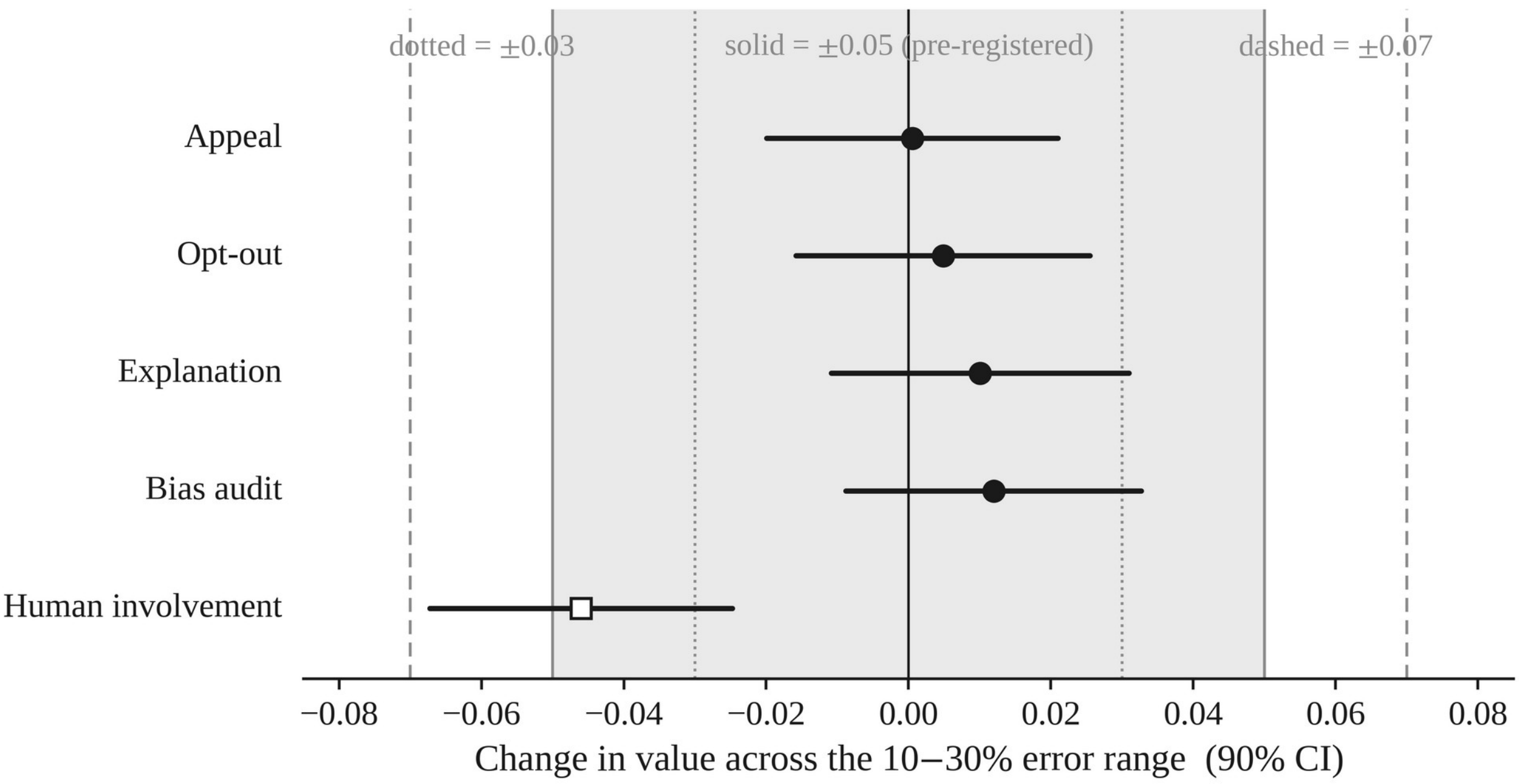

Figure 5: Sensitivity of the equivalence tests to the bound.

Note: Appeal and the opt-out pass the stricter ±0.03 bound; the bias audit and explanation pass the preregistered ±0.05 bound; human involvement passes neither, and only the wider ±0.07 bound (dashed) contains it.

## 5 Discussion

Applicants valued the appeal, the opt-out, and the bias audit as much when the system rarely erred as when it often did, each moderation falling within the preregistered bound. Appeal is the sharpest case, since its corrective function is the most direct. Only human involvement fell outside the bound (Section 4.2). These results do not support the simple error-correction account.

On a performance-independent account, a procedure signals standing and respect, so its value need not depend on what it corrects (Lind & Tyler, 1988; Tyler & Lind, 1992). Folger (1977) found that voice reduced felt injustice even when it did not improve the outcome. Leventhal (1980) treats correctability as one criterion among consistency, accuracy, representativeness, and freedom from bias, so procedures are assessed on several grounds at once. One classic finding runs the other way. Procedural fairness weighs most heavily on people whose outcomes are unfavorable (Brockner & Wiesenfeld, 1996), and that work varies the outcome an individual receives, whereas this design varies the rate at which the system errs.

The human involvement effects are consistent with evidence that people often prefer human decision-makers (Bansak & Paulson, 2024) and that applicants judge fully algorithmic hiring less fair (Lavanchy et al., 2023). They differ from evidence that a supervisory human add-on did not affect fairness perceptions in a lending vignette (Yurrita et al., 2023), plausibly because the present outcome is a forced choice and the decision-authority attribute specifies who decides rather than who supervises.

For employers, performance and procedure answer different demands from job applicants. An employer who adopts a more accurate system while removing the human role or the route of recourse gains on one dimension and loses on another. Applicants may also apply to an employer whose procedures they doubt, so application volume is an imperfect indicator of whether they accept the system as legitimate (Section 4.4).

For system design, these features are not interchangeable. Decision authority settles who makes the call, appeal and the opt-out give the applicant something to invoke, and an audit examines outcomes across cases without offering any individual a route. Adding an audit while removing re-review therefore changes what an applicant can do, not only what an employer can document. A nominal human in the loop does not give the applicant anything to invoke either, since reviewers can defer to the recommendation they are meant to check (Alon-Barkat & Busuioc, 2023; Green, 2022) or degrade its accuracy when they intervene (Sele & Chugunova, 2024). Whether these features also reinforce one another is less clear: the complementarities with human involvement did not hold across scales (Section 4.3), and the design cannot say which combination best protects applicants or improves decision quality.

For regulation, current frameworks assign these functions differently. The EU AI Act requires human oversight of specified recruitment systems (European Parliament & Council of the European Union, 2024, Annex III(4)(a), Arts. 14 and 26(2)). The General Data Protection

Regulation grants a qualified right not to be subject to solely automated decisions, with rights to human intervention and to contest the outcome (European Parliament & Council of the European Union, 2016, Art. 22). New York City's Local Law 144 instead requires a bias audit and a notice explaining how to request an alternative process, which the implementing rules do not oblige an employer to provide (New York City Council, 2021; New York City Department of Consumer and Worker Protection, 2023). Whether an instrument confers a right the applicant can invoke or only system-level oversight therefore matters, because respondents valued the two separately. Preferences do not settle the legal adequacy of any provision, but oversight alone does not exhaust what applicants value.

This study has several limitations. The design measures stated choices in a hypothetical task rather than application behavior, with no status quo or reject-both option (Hainmueller et al., 2015). Respondents were United States job seekers recruited through Prolific, so the sample is not representative of the wider job-seeking population, and the design includes no third-party observer condition. The binary attributes record whether each feature was offered, leaving its quality and implementation unspecified, and the features were presented as costless, whereas real appeal and review involve delay, effort, and uncertain outcomes. The design does not identify why the moderation is small. Applicants may doubt that appeal works as errors mount, or may read its availability as a signal that actual performance exceeds the displayed rate. Finally, the estimates average across respondents, and the eight tasks per respondent give limited individual-level information.

To conclude, performance and procedure carry comparable weight without being interchangeable: cutting wrongful rejections from 30% to 10% is worth about as much as putting a human into the decision, yet the value applicants placed on appeal, the opt-out, and the bias audit did not rise as wrongful rejections became more common. Applicants also put who decides and the routes they themselves can invoke well ahead of a system-level audit. A rule that sets performance standards therefore does not do the work of a rule that confers a right applicants can invoke. What remains open is why procedure holds its value: standing, dignity, and distrust of unaccountable systems all imply the pattern observed here, and telling them apart will require designs that measure these mechanisms directly and that test whether the same pattern holds where the applicant cannot simply take the application elsewhere.

## Declarations

**Preregistration and ethics.** The study received ethical approval from the Department of Methodology at the London School of Economics and Political Science on May 28, 2026. The pre-analysis plan was posted publicly on the Open Science Framework on June 17, 2026 (https://osf.io/5ju4d/), before the data reported here were examined. Deviations from the plan are summarized in Appendix I.

**Participants.** All participants provided informed consent and were compensated at a mean of £0.84, approximately £10.22 per hour.

**Data availability.** Replication data and code are available at https://github.com/chuyao-wang/ai-hiring-replication.

**Funding.** [Insert the funding statement, or state that the research received no specific grant from any funding agency.]

**Acknowledgments.** [Insert acknowledgments.]

**Declaration of competing interests.** The authors declare no competing interests.

## Appendix

This appendix documents the instrument and measures (A), sampling, exclusions, and data quality (B), estimation and full estimates (C), equivalence tests (D), interaction and subgroup estimates (E), further robustness (F), power analysis (G), heterogeneity in the human preference (H), and deviations from the pre-analysis plan (I).

## A Instrument and measures

Beyond the choice and acceptability outcomes, three five-point items measured legitimacy (whether an employer using AI in hiring has the right to decide this way), intention to apply, and intention to withdraw. Intention to withdraw averaged 2.52 (SD = 1.11) and correlated negatively with legitimacy ($r = -0.46$) and with intention to apply ($r = -0.67$). For the legitimacy–application comparison in Section 4.4, the paired difference of 0.389 corresponds to $d_z = 0.42$, $t(1918) = 18.56$.

## B Sampling, exclusions, and data quality

Recruitment followed Section 3.1, with participants from an earlier pilot excluded. The analytic study comprised a soft-launch batch ($n = 150$) and a remaining batch ($n = 1{,}769$) using the same instrument and rules; analysis began on June 27, 2026. A respondent is analytic only if all eight randomization records are present and parse correctly. Withdrawn participants are excluded, an otherwise complete submission with one missing administrative field is retained, and duplicates are removed by identifier. The analytic sample is 1,919. Table B1 reports selected sample characteristics, and Table B2 the attribute-level frequencies.

Table B1: Selected sample characteristics.

| Characteristic | Value |
| --- | --- |
| Respondents (N) | 1,919 |
| Age: mean (SD) | 36.6 (12.1) |
| Age: range | 18–79 |
| Female | 54.6% |
| Male | 44.8% |
| White | 58.6% |
| Black | 16.5% |
| Asian | 9.7% |

| Characteristic | Value |
| --- | --- |
| Mixed | 8.5% |
| Other | 5.1% |
| Unemployed, job-seeking | 35.0% |
| Employed full-time | 31.3% |
| Employed part-time | 18.9% |
| Prior AI evaluation: yes / no / not sure | 74.9% / 12.5% / 12.6% |
| Country of residence: United States | 100% |

Table B2: Attribute-level frequencies across the 30,704 profiles.

| Attribute level | Share of profiles |
| --- | --- |
| Decision authority: AI decides alone | 33.2% |
| Decision authority: AI suggests, human decides | 33.5% |
| Decision authority: AI screens first, human decides | 33.3% |
| Error rate: 10% / 20% / 30% | 33.7% / 32.9% / 33.4% |
| Explanation: available | 49.7% |
| Opt-out: available | 49.9% |
| Appeal: available | 49.5% |
| Bias audit: available | 49.9% |

*Note:* Some Prolific-supplied demographic fields had expired at extraction (at most 1.1% per field); percentages use the full analytic sample as the denominator, and expired and omitted response categories are not displayed. Attribute levels were randomized uniformly, subject to the two profiles in a task differing on at least one attribute.

## C Estimation and full estimates

Section 3.3 gives the primary estimator. Marginal means are reported alongside in Table C3. Decision authority enters as two indicators against the AI-alone reference and is retained as an unordered factor. The error rate enters as a continuous variable in the primary model and as an unordered factor in sensitivity analysis. Marginal means fall in near-equal steps across the three error levels (0.642, 0.499, and 0.357), supporting the linear coding used for the registered moderation test. Table C1 reports the AMCEs.

Table C1: Average marginal component effects on the probability of choice.

| Attribute | AMCE | 95% CI | p |
| --- | --- | --- | --- |
| AI suggests, human decides | +0.287 | [0.273, 0.301] | < 0.001 |
| AI screens first, human decides | +0.257 | [0.244, 0.271] | < 0.001 |

| Attribute | AMCE | 95% CI | p |
|---|---|---|---|
| Appeal (re-review) | +0.156 | [0.145, 0.167] | < 0.001 |
| Opt-out (human from start) | +0.129 | [0.118, 0.140] | < 0.001 |
| Explanation | +0.128 | [0.118, 0.139] | < 0.001 |
| Independent bias audit | +0.068 | [0.058, 0.079] | < 0.001 |
| Error rate (per 1 pp) | −0.014 | [−0.015, −0.014] | < 0.001 |

*Note:* $N$ = 1,919 respondents; 30,704 profile observations. Ordinary least squares linear probability model (LPM); standard errors clustered by respondent. Reference levels as beneath Table 1. The two human-involved levels combine to a pooled human involvement effect of +0.272 (95% CI [0.260, 0.284]), the quantity used in the H1b and H4 contrasts.

Table C2: Planned contrasts for H1b and H4, and the explanation contrast.

| Contrast | Difference | 95% CI | p |
|---|---|---|---|
| Human involvement − appeal | +0.116 | [0.099, 0.133] | < 0.001 |
| Human involvement − opt-out | +0.143 | [0.127, 0.160] | < 0.001 |
| Human involvement − explanation | +0.144 | [0.128, 0.160] | < 0.001 |
| Human involvement − bias audit | +0.204 | [0.188, 0.221] | < 0.001 |

*Note:* Differences compare the pooled human involvement AMCE with appeal, the opt-out, and the bias audit, using respondent-clustered standard errors. These three contrasts implement the preregistered H1b decision rule. The explanation contrast comes from the prespecified separate analysis. The remaining H4 contrasts are appeal − bias audit, +0.088 (95% CI [0.073, 0.103]), and opt-out − bias audit, +0.061 ([0.045, 0.076]), both p < 0.001. N = 1,919 respondents and 30,704 profile observations.

Table C3: Marginal means on the probability of choice.

| Level | Marginal mean | 95% CI |
|---|---|---|
| AI decides alone | 0.320 | [0.311, 0.328] |
| AI suggests, human decides | 0.604 | [0.596, 0.612] |
| AI screens first, human decides | 0.575 | [0.567, 0.583] |
| Error rate 10% | 0.642 | [0.634, 0.651] |
| Error rate 20% | 0.499 | [0.491, 0.507] |
| Error rate 30% | 0.357 | [0.349, 0.366] |
| No explanation | 0.437 | [0.431, 0.443] |
| Explanation | 0.564 | [0.558, 0.570] |
| Opt-out not available | 0.435 | [0.429, 0.441] |
| Opt-out available | 0.565 | [0.559, 0.571] |
| Appeal not available | 0.423 | [0.417, 0.429] |
| Appeal available | 0.579 | [0.573, 0.585] |

| Level | Marginal mean | 95% CI |
|---|---|---|
| Bias audit not available | 0.466 | [0.460, 0.471] |
| Bias audit available | 0.535 | [0.529, 0.540] |

*Note: Marginal mean = probability that a profile with the stated level is chosen, averaging over the other attributes; standard errors clustered by respondent (30,704 profile observations).*

## D Equivalence tests

The equivalence procedure is described in Section 3.3. Appeal, the opt-out, and the bias audit meet the preregistered ±0.05 criterion; human involvement does not. Explanation is shown from its prespecified separate analysis. Figure 5 and Table D1 report sensitivity to alternative bounds.

Table D1: Equivalence tests for error-rate moderation.

| Attribute | Moderation over 10–30% | 90% CI | Equivalent? |
|---|---|---|---|
| Appeal | +0.001 | [−0.020, 0.022] | Yes |
| Opt-out | +0.005 | [−0.016, 0.026] | Yes |
| Explanation | +0.010 | [−0.011, 0.030] | Yes |
| Bias audit | +0.012 | [−0.009, 0.032] | Yes |
| Human involvement | −0.046 | [−0.067, −0.025] | No |

*Note:* Moderation is the change in the attribute's AMCE across the full 10% to 30% error range. Equivalence by two one-sided tests at $\alpha = 0.05$ corresponds to the 90% confidence interval lying within ±0.05. At the stricter ±0.03 bound, appeal and the opt-out remain equivalent; the explanation and bias-audit intervals exceed +0.03 before rounding (upper bounds +0.0304 and +0.0323).

## E Interaction and subgroup estimates

Table E2 reports the full interaction estimates under both links; Section 4.2 gives the substantive treatment of the human-by-error-rate result. Table E1 reports the conditional human involvement effect by subgroup and supports the exploratory analyses in Appendix H. The exploratory joint human-complementarity estimates appear in Table E2, Panel B; the corresponding logit results are weaker.

Table E1: Conditional AMCE of any human involvement by subgroup.

| Subgroup | Conditional AMCE | 95% CI |
|---|---|---|
| All respondents | +0.272 | [0.260, 0.284] |

| Subgroup | Conditional AMCE | 95% CI |
|---|---|---|
| Prior AI experience: yes | +0.269 | [0.255, 0.283] |
| Prior AI experience: no or not sure | +0.282 | [0.258, 0.307] |
| Women | +0.289 | [0.273, 0.304] |
| Men | +0.253 | [0.234, 0.271] |

*Note: The contrast is any human involvement versus AI decides alone. All estimates are exploratory (Appendix H).*

Table E2: Full interaction estimates on the probability of choice, LPM and logit.

| Interaction | LPM estimate | LPM 95% CI | LPM p | Logit coef. (log-odds) | Logit p |
|---|---|---|---|---|---|
| ***Panel A. Error rate × attribute (change in AMCE over the 10–30% error range; logit per 1 pp)*** | | | | | |
| Appeal | +0.001 | [−0.023, 0.025] | 0.94 | −0.001 | 0.87 |
| Opt-out | +0.005 | [−0.020, 0.030] | 0.69 | +0.001 | 0.82 |
| Explanation | +0.010 | [−0.015, 0.034] | 0.45 | +0.002 | 0.64 |
| Bias audit | +0.012 | [−0.013, 0.036] | 0.36 | +0.003 | 0.42 |
| Human involvement | −0.046 | [−0.071, −0.021] | < 0.001 | −0.004 | 0.21 |
| ***Panel B. Procedural feature × human involvement (all four interactions estimated jointly)*** | | | | | |
| Appeal × human | +0.026 | [0.004, 0.047] | 0.020 | +0.056 | 0.31 |
| Opt-out × human | −0.000 | [−0.021, 0.021] | 0.99 | −0.055 | 0.31 |
| Explanation × human | +0.027 | [0.006, 0.048] | 0.011 | +0.081 | 0.14 |
| Bias audit × human | +0.029 | [0.008, 0.051] | 0.008 | +0.117 | 0.036 |

*Note:* LPM interactions use respondent-clustered standard errors. Panel A reports changes in AMCEs across the 10% to 30% error range; Panel B reports differences in AMCEs with versus without human involvement. Logit coefficients are on the log-odds scale. Panel A estimates each error-rate interaction separately; Panel B reports the exploratory joint model; the registered two-interaction model yields the estimates quoted in Section 4.3 (appeal × human +0.025, $p = 0.021$; opt-out × human −0.0003, $p = 0.99$). The human-by-error-rate result is scale-sensitive, and the full complementarity model is exploratory. The appeal, bias-audit, and explanation interactions in Panel B survive a Holm correction on the LPM scale. The p-values of the nonsignificant interactions depend on the finite-sample cluster-robust correction and can differ in the second decimal across software versions; each remains nonsignificant. N = 1,919 respondents and 30,704 profile observations.

## F Further robustness

Each check re-estimates the primary model on a restricted sample or with an added control, and reports how far every attribute effect moves from the Table C1 baseline. Restricting to respondents above a third of the median completion time moves an AMCE by at most 0.001 (n = 1,912); dropping straight-lined ratings by at most 0.003 (n = 1,840); dropping respondents who always chose the same side by at most 0.001 (n = 1,896); and dropping those who failed the manipulation check by at most 0.003 (n = 1,773). Dropping all three quality flags at once moves an AMCE by at most 0.005 (n = 1,687). Controls for task position, for the row position of each attribute, and for profile position, and an unordered coding of the error rate, each move an AMCE by less than 0.001. The derived 20-percentage-point error-rate gain moves by 0.008 under the manipulation-check restriction. Estimates for every specification are reproduced by robustness_appendix_f.py in the replication package.

## G Power analysis

The target of approximately 1,400 respondents was set by the error-rate-moderation and opt-out-by-human tests. Monte Carlo simulations used conservative effect sizes from a separate earlier pilot whose participants were excluded from the analytic study. The analytic-study outcomes were not examined before the plan was posted, and no interim outcome analysis was conducted.

## H Heterogeneity in the human preference

The raw within-respondent difference between choosing human-involved and AI-alone profiles was positive for 77% of the 1,911 respondents exposed to both process types, but with only eight tasks per respondent this statistic is noisy. A simulation holding each respondent's numbers of human-involved and AI-alone profiles fixed produced a standard deviation of 0.27 under a homogeneous-preference benchmark, against an observed 0.30 (Figure H1). The observed dispersion therefore modestly exceeds the fitted benchmark, though the analysis does not identify a model-free individual-level variance.

The prespecified exploratory moderator, prior AI-evaluation experience, did not significantly condition the human involvement AMCE (interaction −0.013, $p$ = 0.36; Table E1). Unregistered analyses found no detectable age interaction and a small gender interaction (women +0.037, $p$ = 0.003); nonsignificant interactions are not treated as evidence of homogeneous preferences.

An exploratory convergence check found only small correlations between the within-respondent human involvement statistic and the legitimacy, application-intention, and withdrawal-intention measures ($|r| \leq 0.14$).

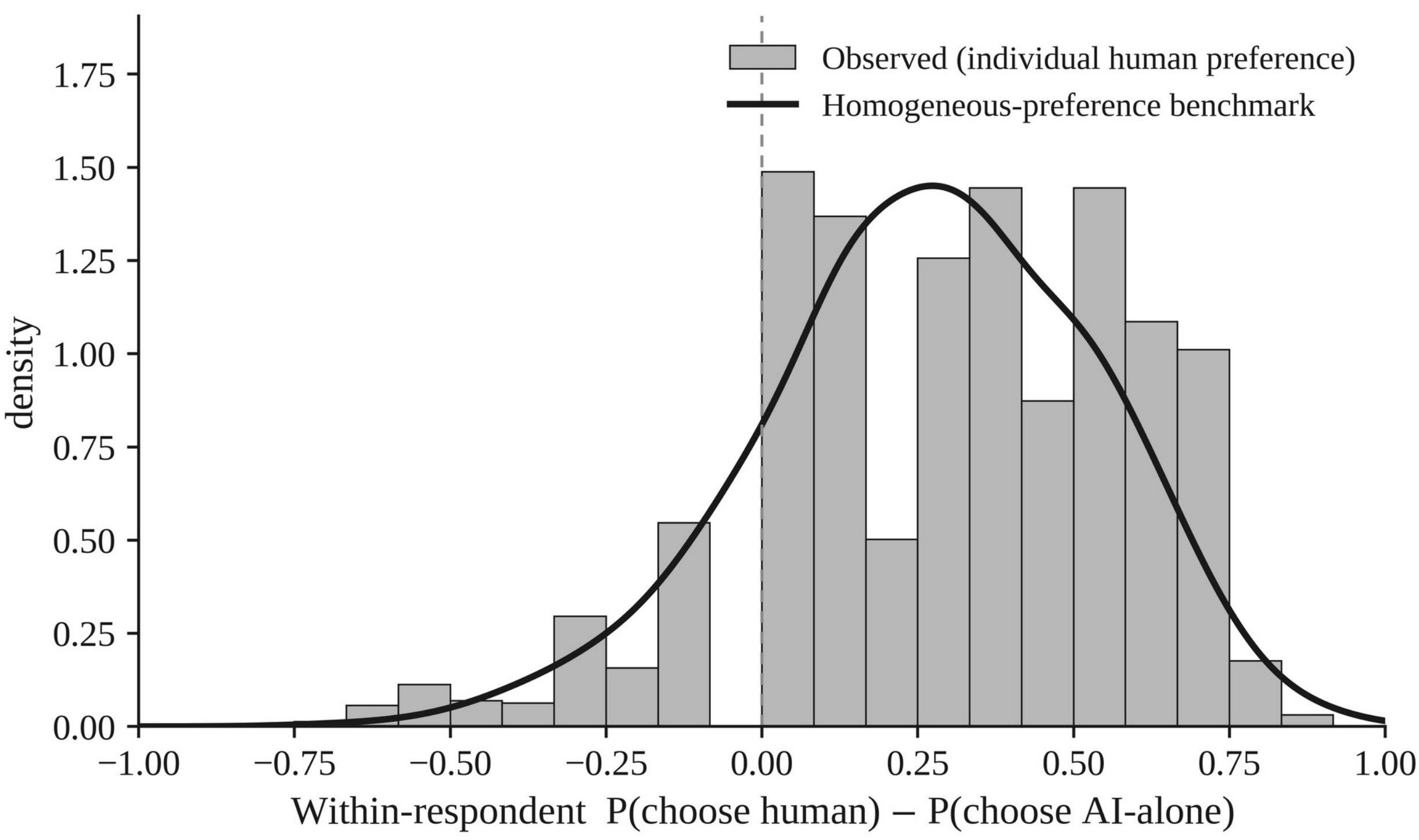

Figure H1: Observed distribution and the homogeneous-preference benchmark.

## I Deviations from the pre-analysis plan

The registered hypotheses and primary estimands are reported as specified in the pre-analysis plan. Table I1 identifies supplementary classifications, post hoc equivalence analyses, alternative interpretations, and unregistered exploratory additions.

Table I1: Pre-analysis plan versus paper, with rationale.

| Item | Pre-analysis plan | This paper | Rationale |
|---|---|---|---|
| **Explanation** | Measured and reported separately; not one of the four confirmatory attributes in H1a, H1b, H2, or H4. | Included as a supplementary non-human feature in figures and equivalence analyses; confirmatory conclusions distinguish it from the registered family. | Keeps the measured feature visible without changing the registered hypotheses. |
| **H2(a) (10%-error condition)** | Registered as H2(a): each feature has a positive AMCE within the 10%-error condition. | Not restated as a separate hypothesis; its 10%-error conditional AMCEs are 0.286 (human involvement), 0.150 (appeal), 0.122 (opt-out), and 0.066 (bias audit). The remaining registered parts, H2(b) and H2(c), are presented as H2a and H2b. | Presentation change only; the distinct conditional estimands and results are preserved. |

| Item | Pre-analysis plan | This paper | Rationale |
|---|---|---|---|
| **Opt-out × human** | Registered as confirmatory but power-limited. | The registered positive interaction was not supported. A post hoc equivalence test is reported descriptively and is not treated as support for H3. | Distinguishes the registered directional test from the supplementary equivalence analysis. |
| **Human × error rate** | Two-sided test of moderation. | Reported as scale-sensitive and not interpreted substantively. | The direction is opposite to H2a, but H2b equivalence is not established for human involvement. |
| **Audit and explanation × human** | Only appeal and opt-out interactions with human involvement registered (H3). | All four procedural-feature × human interactions estimated. | Exploratory extension; Holm-adjusted; flagged as exploratory in Section 4.3. |
| **Age and gender** | Prior AI experience was prespecified as exploratory; age and gender were not registered. | Age and gender examined as additional moderators. | Exploratory and unregistered; flagged in Appendix H. |
| **Heterogeneity simulation** | Not prespecified. | Homogeneous-preference simulation used as a model-based benchmark for observed dispersion. | Reported as exploratory without interpreting the variance difference as a model-free true effect. |
| **Decision-authority sequence × error rate** | Not prespecified. | The “AI suggests” versus “AI screens first” contrast is interacted with the error rate and reported in Section 4.2. | Exploratory and unregistered; flagged as such in the text. |
| **Framing of H2** | Instrumental vs. procedural-justice (non-instrumental) valuation. | Simple error-correction vs. performance-independent process account. | Relabeling of the same registered constructs; estimands, models, and decision rules unchanged. |
| **Wording of H2a and H2b** | Hypotheses stated in terms of the wrongful-rejection rate. | Hypotheses stated in terms of the error rate; its operational definition (the share of qualified applicants wrongly rejected) is given in Table 1. | Terminology alignment for readability; the construct, estimand, and decision rule are unchanged. H2a states the moderation claim alone; the diagnostic priority of appeal is stated in the text rather than within the hypothesis. |
| **H1** | Registered as one hypothesis combining the positive effects and the ordering of human involvement. | Stated as H1a (positive effects) and H1b (ordering). | Wording only; the estimands, models, and decision rules are unchanged. |

| Item | Pre-analysis plan | This paper | Rationale |
|---|---|---|---|
| **Wording of H5** | Registered as a within-respondent comparison of intention to apply and belief in legitimacy. | Reported descriptively in Section 4.4; not stated as a hypothesis and not used to claim that applicants apply despite disapproval. | The comparison rests on two single items measuring distinct constructs and addresses a question separate from the conjoint design; the registered estimand, paired test, and result are unchanged. |
| **Sample size** | Target N ≈ 1,400 valid completions plus a buffer. | 1,919 analytic respondents. | Recruitment exceeded the target as the funding allowed. |